\documentstyle[12pt]{article} 
\begin{document}


\newcommand\mathC{\mkern1mu\raise2.2pt\hbox{$\scriptscriptstyle|$}
                {\mkern-7mu\rm C}}		       
\newcommand{\mathR}{{\rm I\! R}}                


\begin{titlepage}

\hspace{8truecm} Newton Institute NI97036

\begin{center}
        {\large\bf On multi-particle entanglement}
\end{center}
\vspace{1 truecm}
\begin{center}
        N.~Linden${}^{a,}$\footnote{email: n.linden@newton.cam.ac.uk} and
S.~Popescu${}^{a,b,}$\footnote{email: s.popescu@newton.cam.ac.uk}\\[0.4cm]
       ${}^a$ Isaac Newton Institute for Mathematical Sciences\\
        20 Clarkson Road\\
        Cambridge CB3 0EH\\ 
        United Kingdom\\
 \end{center}
\bigskip
\begin{center}
        ${}^b$BRIMS Hewlett-Packard Labs.\\
        Filton Road, Stoke Gifford\\
        Bristol, BS12 6QZ\\
        United Kingdom\\
\end{center} 
\medskip 

\begin{center} October 1997\end{center}
 
\begin{abstract}
We build, using group-theoretic methods, a general framework for approaching
multi-particle entanglement.  As far as entanglement is concerned, two states of
$n$ spin-1/2 particles are equivalent if they are on the same orbit of the group
of local rotations ($U(2)^n$).  We give a method for finding the number of
parameters needed to describe inequivalent $n$ spin-1/2 particles states.  We
also describe how entanglement of states on a given orbit may be characterized by
the stability group of the action of the group of local rotations on any point
on the orbit.
\end{abstract}
 
\end{titlepage}
\section{Introduction}

\bigskip

Discovered in 1964 by J.  Bell \cite{Bell}, the existence of non-local 
correlations
among remote quantum systems is one of the most fascinating quantum phenomena.
But while for long time these correlations were considered more as a curiosity, 
recently they have found a large range of applications, forming the very bases 
of 
quantum communication and quantum computation; obviously the interest in better 
understanding these correlations has increased dramatically.

Traditionally, starting with Bell, the example which has been most studied was 
that of 
non-local correlations between {\it two} remote quantum particles. However, it 
is now clear that the correlations among more than two remote particles present 
novel and highly nontrivial aspects compared to two-particles entanglement. (See 
for example the correlations generated by the GHZ state \cite{GHZ}.) 
Nevertheless, 
very little is known yet about multi-particle entanglement. It is the aim of the 
present paper to take a few steps towards understanding the general structure of 
multi-particle entanglement.

The key element in our approach is to note that two states which can be 
transformed one into another by {\it local} operations (unitary transformations) 
are equivalent as far as their non-local properties are concerned. This leads us 
to investigate the properties of the Hilbert space of $n$ spin-1/2 
particles under local unitary transformations.

We find the following picture emerging.
\begin{itemize}
\item
Each particular state $\Psi$ belongs to an equivalence class comprised of by all 
states which can be obtained from $\Psi$ by acting on it with local unitary 
operators; all states in a class are equivalent as far as non-locality is 
concerned.  Obviously, the Hilbert space of states decomposes completely into 
equivalence classes, or ``orbits", under the action of the group of local 
unitary 
transformations.
\item
An arbitrary state $\Psi$ of $n$ spin 1/2 particles is described by $2^n$ 
complex parameters. Some of these parameters (or functions of them) specify the 
equivalence class to which $\Psi$ belongs. These parameters (or functions) are 
obviously {\it invariants} under local transformations. The reminder describe 
where $\Psi$ is situated inside the equivalence class - they do change under 
local transformations.

Incidentally, two-particles entanglement is technically so much simpler to study
than multi-particles entanglement because there is a simple way to identify the
invariant parameters - the Schmidt decomposition.  Indeed, let $e_i\otimes e_j$, 
$i,j=1,2$
be some arbitrary base vectors in the Hilbert space of the two particles then a
general state of two particles is given by
\begin{equation}
\Psi=\sum_{i,j}\alpha_{ij}e_i\otimes e_j.
\end{equation}
However, by choosing some appropriate base vectors for each particle, the double
sum in (1) can be reduced to a single sum
\begin{equation}
\Psi=\sum_i \beta_i{f_i\otimes f_i}.
\end{equation}
The Schmidt coefficients are manifestly invariant under local transformations. 
Indeed, local unitary transformations can only change the 
Schmidt base vectors, but not the Schmidt coefficients.

\begin{equation}
\Psi=\sum_i \beta_i{f_i\otimes f_i}\rightarrow \Psi'=\sum_i 
\beta_i{f'_i\otimes f'_i}.
\end{equation}

As it is well-known, for multi-particle states in general there exists no
similar simple decomposition \cite{Asher}.  What can one than do?  Given our
above analysis of multi-particle entanglement, it is now clear that instead of
simply trying to find something which {\it formally} resembles the Schmidt
decomposition, we should try to follow its spirit, not its form. That is, to try
and find a representation which separates local and non-local parameters.

As an important result, we find that for large $n$, most of the parameters 
describe non-local properties. This is opposite to the case of small $n$ - for 
two spins, out of the 8 real parameters which describe  a generic (unnormalized) 
state, only 1, the unique independent Schmidt coefficient, has non-local 
significance.

\item Finally, we note that in the case of two-particle entanglement some of the 
states are, in some sense, special. Such states are the direct-products and the 
singlet-like states. We show that the special nature of these states is 
determined by their invariance properties. Namely, for these special states 
there are more local actions which leave them unchanged than 
in the case of generic states. For example in the case of a singlet
\begin{equation}
\Psi={1\over{\sqrt2}}(e_1\otimes e_2-e_2\otimes e_1)
\end{equation}
where $e_1$ and $e_2$  represent spin polarized ``up" or ``down" 
say, along the $z$ axis, identical rotations of the two spins leave the state 
unchanged.

Furthermore, such enhanced invariance properties are in fact common for all 
states in an equivalence class, and thus characterize the class itself.
To find the ``special" equivalence classes, we have therefore to study their 
invariance properties. We argue that these ``special" classes describe 
fundamentally different types of entanglement while a generic class represents a 
combination of different types of entanglement.

\end{itemize}

Group-theoretically, the situation is the following. The space of states of $n$ 
spin 1/2 particles is the $n$-fold tensor product 
$\mathC^{2^n}=\mathC^2\otimes...\otimes \mathC^2$, and the group of local 
transformations is the 
$n$-fold product $U(2)^n=U(2)\times...\times U(2)$, (each copy of $U(2)$ acting 
on a different spin, i.e. on the corresponding copy of $\mathC^2$). The 
equivalence 
classes are {\it orbits} under the action of the local transformations group. 
Hence, the space of orbits is 
\begin{equation}
{{\mathC^{2^n}}\over{U(2)\times...\times U(2)}};
\end{equation}
this 
is the main mathematical object we are investigating.

The number of parameters needed to describe the position of $\Psi$ on its orbit 
is the {\it dimension} of the orbit. Not all orbits have the same dimension. As 
noted above, there are ``special" orbits - singular orbits- which have higher 
invariance, i.e. lower dimension.

The total number of parameters ($2^n$ complex parameters = $2^{n+1}$ real 
parameters) describing the space of states minus the number of parameters 
describing a generic orbit (the dimension of the orbit), gives the number of 
parameters describing the location of the orbit in the space of orbits, i.e. the 
number of parameters describing the non-local properties of the states.

\section{The number of parameters needed to describe inequivalent states}

In this section we are interested in finding out how many parameters are needed
to describe the space of orbits of the action of $U(2)^n$ on the space of
states, i.e.  the number of parameters which describes inequivalent states.  To
do this it will be convenient to find the (real) dimension of a general orbit;
the number of parameters is then found by subtracting this number from
$2^{n+1}$.

A lower bound on this number can be obtained by a simple argument of counting 
parameters. Each of the $n$ copies of the local unitary group $U(2)$ is 
described by 4 real parameters. Thus there can be no more than $4n$ parameters 
describing local properties of the states, and hence at least $2^{n+1}-4n$ 
non-local parameters (i.e., invariants under local transformations)\footnote{For 
convenience we always consider non-normalized states, and thus the norm also 
appears as one of the invariant parameters.}.

One can immediately see that for large $n$ almost all parameters have non-local 
significance.

The above bound is, in general, not satisfied. The reason is that not all $4n$ 
parameters describing the local transformations lead to independent effects. For 
example, equally changing the phase of all states of any particular spin has the 
same effect as changing the phases of any other. Hence, at least, the group of 
local transformations reduces from $U(2)^n$ to $U(1)\times SU(2)^n$ which has 
dimension $3n+1$. This leads to a better lower bound on the number of non-local 
parameters of $2^{n+1}-(3n+1)$.

This is, however, not the end of the story. We will find below that the number 
of parameters describing independent local transformations may be fewer (and 
correspondingly, the number of non-local parameters larger).

\bigskip
\subsection{Dimension of a general orbit}
\bigskip

 To find
the dimension of a general orbit it is simplest to work infinitesimally.
Thus, in general, associated to the action of each element of a Lie algebra of a
Lie group $K$ which acts on a space $V$ there is a vector
field: take an  element $T$ of a basis for the Lie algebra,  the action of the
group element $k=\exp i\epsilon T\in K$ on an element $v\in V$ induces an
action on functions from $V$ to $\mathC$; and the vector field, $X_T$, 
associated
to the Lie algebra element $T$ is found by differentiating:

\begin{equation}
X_T f(v) {\buildrel {\rm def}\over =} {\partial\over\partial\epsilon}
f(e^{i\epsilon T}v)\vert_{\epsilon =0}.
\label{field}
\end{equation}

The linear span of vector fields at the point $v$ associated with the whole Lie
algebra forms the tangent space to the orbit at the point $v$ and so the number
of linearly independent vector fields at this point gives the dimension of the
orbit.

\bigskip
\subsection{A single spin}
The case $n=1$ helps to illustrate the general formalism.  The space of states
has real dimension four (complex dimension two).  It is also clear that the
action of a unitary operator on a vector cannot change its norm, so that the
dimension of the space of orbits must be at least one (in fact we will soon see 
that it
is precisely 
one).   However, the group $U(2)$ has dimension four so that the set of vector 
fields
associated to an arbitrary basis for the Lie algebra cannot be linearly
independent.

In the representation of $U(2)$ acting on $\mathC^2$ a convenient Hermitian
basis for the Lie algebra is
\begin{equation}
\sigma_x = 
\left(\begin{array}{cc} 
0& 1 \\
1&0
\end{array}\right)
\quad
\sigma_y = 
\left(\begin{array}{cc} 
0& -i \\
i&0
\end{array}\right)
\quad
\sigma_z = 
\left(\begin{array}{cc} 
1& 0 \\
0&-1
\end{array}\right)
\quad
1_2 = 
\left(\begin{array}{cc} 
1& 0 \\
0&1
\end{array}\right)
.
\end{equation}

Now take an element 
\begin{equation}
\Psi=
\left(\begin{array}{l} 
\alpha \\
\beta
\end{array}\right) \in \mathC^2
\end{equation}
and consider the infinitesimal change under a transformation in the direction
$\sigma_x$:
\begin{equation}
\delta \Psi= i\epsilon \sigma_x \Psi 
= i\epsilon 
\left(\begin{array}{cc} 
0& 1 \\
1&0
\end{array}\right) 
\Psi
= 
\left(\begin{array}{l} 
i\epsilon\beta \\
i\epsilon\alpha
\end{array}\right).
\end{equation}
So that under a  group transformation close to the identity,
\begin{equation}
\Psi=
\left(\begin{array}{l} 
\alpha \\
\beta
\end{array}\right) 
\mapsto
\Psi+\delta \Psi
=
\left(\begin{array}{l} 
\alpha + i\epsilon\beta \\
\beta + i\epsilon\alpha
\end{array}\right).
\end{equation}
We now write everything in terms of real variables:
\begin{equation}
\alpha = c_1 + i d_1;\quad \beta = c_2 + i d_2.
\end{equation}
so that
\begin{equation}
\Psi=
\left(\begin{array}{l} 
c_1 \\
d_1\\
c_2\\
d_2
\end{array}\right) 
\quad\hbox{\rm and}\quad 
\delta \Psi = \epsilon
\left(\begin{array}{l} 
-d_2\\
c_2\\
-d_1\\
c_1
\end{array}\right) 
=\epsilon
\left(\begin{array}{cccc} 
0&0&0&-1\\
0&0&1&0\\
0&-1&0&0\\
1&0&0&0
\end{array}\right) 
\left(\begin{array}{l} 
c_1 \\
d_1\\
c_2\\
d_2
\end{array}\right).
\end{equation}
Thus there is an induced action on a function $f(v) = f(c_1,d_1,c_2,d_2)$:
\begin{equation}
f(c_1,d_1,c_2,d_2) \mapsto f(c_1-\epsilon d_2,d_1 +\epsilon
c_2,c_2-\epsilon d_1,d_2 + \epsilon c_1) .
\end{equation} 
Differentiating with respect to $\epsilon$ we find: 
\begin{equation}
{\partial f\over \partial \epsilon}\Bigg\vert_{\epsilon=0} =
\left(
-d_2 {\partial \over \partial c_1} + c_2 {\partial \over \partial d_1}
- d_1 {\partial \over \partial c_2} + c_1 {\partial \over \partial d_2}
\right) f.
\end{equation}
We write the vector field associated to this Lie algebra element $\sigma_x$ as
\begin{equation}
\left(
-d_2 {\partial \over \partial c_1} + c_2 {\partial \over \partial d_1}
- d_1 {\partial \over \partial c_2} + c_1 {\partial \over \partial d_2}
\right) = u_x.\nabla 
\quad\hbox{where}\quad u_x = 
\left(\begin{array}{l} 
-d_2\\
c_2\\
-d_1\\
c_1
\end{array}\right).
 \end{equation}
In a similar way we may find the vectors $u_y, u_z$ and $u_1$ associated to
transformations by $\sigma_y, \sigma_z$ and $1_2$:

\begin{equation}
u_y = 
\left(\begin{array}{l} 
c_2\\
d_2\\
-c_1\\
-d_1
\end{array}\right)\quad 
u_z = 
\left(\begin{array}{l} 
-d_1\\
c_1\\
d_2\\
-c_2
\end{array}\right)
\quad
u_1 =
\left(\begin{array}{l} 
-d_1\\
c_1\\
-d_2\\
c_2
\end{array}\right).
\end{equation}
It is not too difficult to check that only three of these four vectors are
linearly independent.  Indeed
\begin{equation}
2(d_1d_2 +c_1c_2)u_x + 2(c_1d_2 -d_1c_2)u_y + (c_1^2 + d_1^2 -c_2^2 -d_2^2)u_z
-(c_1^2 + d_1^2 +c_2^2 +d_2^2)u_1 = 0.
\end{equation}
Thus the dimension of the orbit is three and so there is one parameter (the
norm) which describes the different orbits.
\bigskip
\subsection{Two spins}
In a similar way we may analyze the case of two spins.  A general vector may be
written
\begin{equation}
\Psi = \sum_{i,j=1}^2 \alpha_{ij} e_i\otimes e_j 
= \sum_{i,j=1}^2 (c_{ij}+id_{ij}) e_i\otimes e_j,
\end{equation}
where $\{e_1,e_2\}$ is a general basis of $\mathC^2$. In the representation of 
$U(2)^2$ on $\mathC^4$
we may use the following basis for the eight Lie algebra elements: 
\begin{equation}
\sigma_x\otimes 1_2,\ \sigma_y\otimes 1_2,\ \sigma_z\otimes 1_2,\ 
1_2\otimes 1_2,1_2\otimes\sigma_x,\ 1_2\otimes\sigma_y,\ 1_2\otimes\sigma_z,\ 
1_2\otimes 1_2.
\end{equation}
One sees that the element $1_2\otimes 1_2$ appears twice, so that in fact there
are only seven different Lie algebra elements to consider.  If we choose the
following order for the coordinates of the eight dimensional real vector space:
\begin{equation}
(c_{11},d_{11},c_{12},d_{12},c_{21},d_{21},c_{22},d_{22}),
\end{equation}
then the derivative operator is
\begin{equation}
\nabla_8 =
(
{\partial\over\partial c_{11}},
{\partial\over\partial d_{11}},
{\partial\over\partial c_{12}},
{\partial\over\partial d_{12}},
{\partial\over\partial c_{21}},
{\partial\over\partial d_{21}},
{\partial\over\partial c_{22}},
{\partial\over\partial d_{22}}
),
\end{equation}
and the vector fields are all of the form $u.\nabla_8$.  
The vectors
 $\{ u^{(1)}_{x},u^{(1)}_{y},u^{(1)}_{z},u_{\rm one},\hfil\break
 u^{(2)}_{x},u^{(2)}_{y},u^{(2)}_{z} \}$ associated to the Lie algebra elements
$\{\sigma_x\otimes 1_2,\ \sigma_y\otimes 1_2,\hfil\break \sigma_z\otimes 1_2,\ 
1_2\otimes 1_2,1_2\otimes\sigma_x,\ 1_2\otimes\sigma_y,\ 1_2\otimes\sigma_z\}$
respectively (the superscript on $u$ refers the component in the tensor product,
the subscript the Lie algebra element) are
\begin{eqnarray}
u^{(1)}_x =
(-d_{21},c_{21},-d_{22},c_{22},-d_{11},c_{11},-d_{12},c_{12})^T,\nonumber\\
u^{(1)}_{y} =
(c_{21},d_{21},c_{22},d_{22},-c_{11},-d_{11},-c_{12},-d_{12})^T,\nonumber\\
u^{(1)}_{z} =
(-d_{11},c_{11},-d_{12},c_{12},d_{21},-c_{21},d_{22},-c_{22})^T,\nonumber\\
u_{\rm one} =
(-d_{11},c_{11},-d_{12},c_{12},-d_{21},c_{21},-d_{22},c_{22})^T,\nonumber\\
u^{(2)}_{x} =
(-d_{12},c_{12},-d_{11},c_{11},-d_{22},c_{22},-d_{21},c_{21})^T,\nonumber\\
u^{(2)}_{y} =
(c_{12},d_{12},-c_{11},-d_{11},c_{22},d_{22},-c_{21},-d_{21})^T,\nonumber\\
u^{(2)}_{z} =
(-d_{11},c_{11},d_{12},-c_{12},-d_{21},c_{21},d_{22},-c_{22})^T.
\label{uvectors}
\end{eqnarray}
It may be shown that only six of these vectors are linearly independent for 
general values of the $c_{ij}$ and $d_{ij}$.  Thus the dimension of the generic 
orbit is
six and therefore the number of parameters  describing  the different orbits is 
two.  This confirms
the well-known result that any state of two spins is equivalent, under local 
rotations, to one of
the  form
\begin{equation}
N(\cos\phi\ e_1\otimes e_1\ +\ \sin\phi\  e_2\otimes e_2).
\label{schmidt}
\end{equation}

\subsection{Three spins}
\bigskip
A computation similar to the one in the above subsections shows that in the case 
of 3 spin 1/2 particles the dimension of a generic orbit is 10, and hence the 
number of real non-local parameters (including the norm) is 6 ($=2^{3+1}-10$).

It is interesting to note that in this case {\it all} the $3\times3+1=10$ 
parameters describing the local transformations $U(1)\times SU(2)^3$ are 
actually independent. 

By brute force one can show that any 3 spin 1/2 particle state is equivalent, up 
to local transformations to \footnote{This result was found independently by J. 
Schlienz \cite{sch}}

\begin{eqnarray}
N\cos\alpha e_1\otimes(\cos\beta e_1\otimes e_1+\sin\beta e_2\otimes 
e_2)+\nonumber\\ 
N\sin\alpha \cos\gamma e_2\otimes(\sin\beta e_1\otimes e_1-\cos\beta e_2\otimes 
e_2)+\nonumber\\
N\sin\alpha \sin\gamma e_2\otimes(\cos\delta e_1\otimes e_2+e^{i\eta}\sin\delta 
e_2\otimes 
e_1).
\end{eqnarray}

A systematic way of finding the invariants is given in the next section.

 \section{Invariants} For some purposes one might wish to
know whether or not two states are on the same orbit, i.e. are equivalent.  In 
principle one can
take the ideas of the previous section further to find invariants of the orbits. 
 For consider any
function on the space of states.  If it is invariant under the action of the 
group then in
particular it is invariant under infinitesimal group transformations.  Thus it 
must be annihilated
by the vector fields associated to the infinitesimal group  transformations.  
Therefore in order to
find a set of infinitesimal invariants one has to solve a set of simultaneous 
partial differential
equations; the number of such equations is the number of linearly independent 
vectors associated
with the Lie algebra, as in the previous section.  

If we label the Lie algebra elements of local transformations $\{T_i\}$, 
$i=1...3n+1$ (corresponding to the local transformations group $U(1)\times 
SU(2)^n$, see section 2) then the vector fields $X_{T_i}$ are derived as in eq. 
(\ref{field}) and an invariant function satisfies

\begin{equation}
X_{T_i}f=0, ~~~~i=1...3n+1,
\end{equation}

a set of $3n+1$ simultaneous linear partial differential equations. 
The method of characteristics
allows one to solve the problem in principle, subject to being able to perform
the integrals which arise.   Unfortunately, one can easily see that the problem
becomes very difficult, even for two spins, for in this case one has to solve 
six
simultaneous partial differential equations\footnote{As we saw in section 2.3 
only six of the seven vector fields are linear independent in this case.}.

It may turn out to be more profitable to realize that one can write down a
series of {\it polynomial} expressions which are manifestly invariant under the
local actions. We will first show a few examples and then discuss the general 
case.

\subsection{Examples}
\bigskip
In the
case of one spin, with general state
\begin{equation}
\Psi= \sum_{i=1}^2\alpha_i e_i,
\end{equation}
one can easily see that the expression
\begin{equation}
\sum_{i=1}^2\alpha_i \alpha_i^*
\end{equation}
(i.e. the norm of the state) is invariant under local unitary transformations.

In the case of two spins, with general state $\Psi=\sum_{i,j=1}^2\alpha_{ij}
e_i\otimes e_j$, the norm of the state is invariant and given by a similar
expression: 
\begin{equation}
I_1 = \sum_{i,i_1,j,j_1=1}^2\alpha_{ij} 
\alpha_{i_1j_1}^*\delta_{ii_1}\delta_{jj_1}\quad =
\quad\sum_{i,j=1}^2\alpha_{ij} \alpha_{ij}^*. 
\end{equation}
There is, however a second, quartic, expression which is functionally 
independent
of $I_1$ which is also clearly invariant, since the indices have been
contracted with the invariant tensor $\delta$:
\begin{eqnarray}
I_2 = \sum_{1}^2\alpha_{ik} \alpha_{i_1m}^*\alpha_{jm_1} \alpha_{j_1k_1}^* 
\delta_{ii_1}\delta_{jj_1}\delta_{kk_1}\delta_{mm_1}\nonumber\\
 = 
\sum_{1}^2\alpha_{ik} \alpha_{im}^*\alpha_{jm} \alpha_{jk}^* 
= {\rm Trace}\left(\left(\alpha\alpha^\dagger\right)^2\right).
 \end{eqnarray}

 In the familiar form of the Schmidt coefficients eq. (\ref{schmidt})
 
 \begin{eqnarray}
 I_1=N^2\nonumber\\
 I_2=N^4(\cos^4\phi+\sin^4\phi).
 \end{eqnarray}
Since we know  that in the case of two
spins there can only be two invariants, any further invariants must be able to 
be
written in terms of $I_1$ and $I_2$.  For example, consider 
\begin{equation}
I_3 = \sum_{i,j,k,m,n,p=1}^2\alpha_{ik} \alpha_{im}^*\alpha_{jn} \alpha_{jk}^*
\alpha_{pm}\alpha_{pn}^* = {\rm 
Trace}\left(\left(\alpha\alpha^\dagger\right)^3\right).
\end{equation}
By noting, for example, that  the $2\times 2$ matrix $\alpha\alpha^\dagger$ is
hermitian and satisfies a quadratic equation (by the Cayley Hamilton theorem), 
one
may show that \begin{equation}
I_3 = {1\over 2}\left(3I_1 I_2 - I_1^3\right).
\end{equation}
In a similar way one may see that all higher order invariants are of the form
\begin{equation}
I_N = {\rm Trace}\left(\left(\alpha\alpha^\dagger\right)^N\right), \quad N\geq 3
\end{equation}
and are expressible in terms of $I_1$ and $I_2$. 

\subsection{General case}
\bigskip

A generic state of $n$ spin 1/2 particles can be written as
$$\Psi=\sum_{i_1,i_2,..i_n=1}^2 \alpha_{i_1i_2...i_n}e_{i_1}\otimes 
e_{i_2}\otimes...\otimes e_{i_n}.$$
Then a general polynomial expression in the coefficients is 
\begin{equation}
\sum 
c_{i_1....k_n...}\alpha_{i_1i_2...i_n}\alpha_{j_1j_2...j_n}...\alpha^*_{k_1k_2..
.k_
n}.....
\label{poly}
\end{equation}
If the polynomial (\ref{poly}) has equal numbers of $\alpha$ and $\alpha^*$ and 
all the 
indexes of $\alpha$ are contracted with those of $\alpha^*$, each index being 
contracted with an index located on the same slot (i.e. if $ c_{i_1....k_n...}$ 
are appropriate products of $\delta$'s) then the polynomial is manifestly 
invariant.

For example, in the
case of three spins with generic state 
\begin{equation}
\Psi=\sum_{i,j,k=1}^2\alpha_{ijk}e_i\otimes e_j\otimes e_k,
\end{equation}
there is one quadratic invariant, the norm, there are the quartic invariants 
(in addition to
the square of the norm),
 \begin{eqnarray}
J_1 = \sum_{1}^2\alpha_{ijk} \alpha_{ijm}^*\alpha_{pqm} 
\alpha_{pqk}^*\nonumber\\
J_2=\sum_{1}^2\alpha_{ikj} \alpha_{imj}^*\alpha_{pmq} \alpha_{pkq}^*\nonumber\\
J_3=\sum_{1}^2\alpha_{kij} \alpha_{mij}^*\alpha_{mpq} \alpha_{kpq}^*,
\end{eqnarray}
and so on, the different invariants arising by contracting indices in different 
ways.

Furthermore, one can prove that {\it all} invariant polynomials are constructed 
in this way. The proof of this theorem (not given here) is based on the fact 
that all polynomial functions of $k$ vectors in $\mathC^2$, invariant under 
$U(2)$ 
are polynomials in the inner-product of the vectors\cite{Spivack}.

A key issue is the following. There are infinitely many polynomial invariants. 
We need  to be able to construct from these polynomials a complete set of 
functionally
independent invariants for arbitrary numbers of spins.  Fortunately, given any 
set of polynomials,
there is a algorithmic procedure of determining the relations between them using 
the theory of
Grobner bases \cite{Grobner}.  Thus there is a systematic way of constructing 
sufficient independent
invariants to classify states; indeed the procedure applies to more general 
groups and
representations than $U(2)$ acting on $\mathC^2$.  Firstly determine the number 
$N_I$ of
independent invariants using the ideas of the previous section.  Then determine 
the number of
independent invariants at lowest order (in the case of $U(2)^n$ acting on 
$\otimes^n\mathC^2$ there
was just one, the norm, of the form $\alpha\alpha^*$).  If this number is less 
than $N_I$,
construct the invariants at the next order and see which of these are 
functionally independent of
each other and the ones previously constructed.  The procedure continues until 
$N_I$ are found. 
There is a simple formula, the Molien formula, for the generating function of 
the number of
linearly independent invariants at each order which may well simplify the 
task\cite{Grobner}.

\section{Orbit Types}
As discussed in the introduction, a further important question that the group 
theoretic approach allows one to
address is what types of entanglement can occur.  One can do this by recalling 
that
by definition any group $G$ acts transitively on an orbit $O$ and thus an orbit 
may be
written as
\begin{equation}
O = G/H
\end{equation}
where $H$ is the stability group of any point on the orbit.  Thus the space of 
states
of $n$-spins, $\mathC^{2^n}$ breaks up into orbits each of which is 
characterized by
its stability group.  Each stability group is a subgroup of $U(2)^n$, so the
issue is then to find which subgroups occur as stability groups.  A generic 
orbit
will have a certain stability group, but there are also special cases are where 
an
orbit has a larger symmetry group.  If we denote by $H_{\Psi}$ the invariance 
group 
of the
state $\Psi$, we will see that states with ``maximal"
symmetry are particularly interesting.  By states of ``maximal" symmetry, we 
mean 
those
states $\Psi$ for which there are no others which have an invariance group which 
contain 
$H_{\Psi}$ as a proper subgroup.

One systematic way to analyze the space of states, in principle, is to use the
infinitesimal methods of section 2.  Consider the 2-spin case.  We found that of 
the
eight generators of $U(2)^2$, only six were linearly independent for generic 
states so
that generic orbits have a two dimensional invariance group.  However there will 
be
some values of the parameters describing the states for which the number of 
linearly
independent vectors is smaller than six.  Finding these points is a problem in 
linear
algebra.  Unfortunately the complexity of the calculation seems to make it
impractical.

An alternative approach is to make use of the fact that every stability group is
a subgroup of $U(2)^n$.  One can make a list of subgroups of $U(2)^n$  and check
which subgroups occur as stability groups.  Goursat's theorem \cite{AlperinBell}
gives a complete characterization of subgroups of any direct product of two 
groups
and this enables one, in principle, to produce this list.  The complete set of
subgroups, even of $ U(2)\times U(2)$ is considerable, once all discrete 
subgroups
are taken into account.  However, the example below shows that much progress in
understanding the space of states can be made by considering only continuous 
subgroups
in the first instance.

As an example, consider a  (fairly general) three-spin state of the form
\begin{equation}
\Psi = a e_1\otimes e_1\otimes e_1 + be_2\otimes e_2\otimes e_2 + 
c e_1\otimes e_1\otimes e_2 + d e_2\otimes e_1\otimes e_1.
\end{equation}

In order to find whether this state is invariant under any continuous 
(connected)
group, it suffices to check whether it is annihilated by any Lie algebra 
element.  As mentioned in section 2, since each copy of $U(2)$ in the
group $U(2)^3$ contains a $U(1)$ subgroup corresponding to changing
the global phase of the state, it suffices to consider $SU(2)^3\times U(1)$;
thus the phase is counted only once.
The most general Lie algebra element in this case is
\begin{eqnarray}
T &= \alpha_1(\sigma_x)_1 +  \alpha_2(\sigma_x)_2 + \alpha_3(\sigma_x)_3
 + \beta_1(\sigma_y)_1 + \beta_2(\sigma_y)_2 + \beta_3(\sigma_y)_3\nonumber\\
&\qquad + \gamma_1(\sigma_z)_1 + \gamma_2(\sigma_z)_2 + \gamma_3(\sigma_z)_3
 + \delta 1_8,
\end{eqnarray}
where 
\begin{equation}
(\sigma_x)_1 = \sigma_x\otimes 1_2 \otimes 1_2;\quad 
(\sigma_x)_2 = 1_2 \otimes\sigma_x\otimes 1_2 \quad\hbox{etc.}
\end{equation}
and $1_8$ is the identity element 
\begin{equation}
 1_8 =
1_2 \otimes 1_2\otimes 1_2. 
\end{equation}
By direct calculation one can check that if $a,b,c$ and $d$ are all non-zero, 
then
the state is not annihilated by any non-zero Lie algebra element so that the 
state is
not invariant under any continuous (connected) group.  

The special cases, where the state does have an invariance group, are 
interesting,
however: consider first the case $a=0$.  If $b,c$ and $d$ are all non-zero then 
we
find that the state is annihilated by the Lie algebra element with $\gamma_1=
-\gamma_2=\gamma_3=\delta$ with all other coefficients in $T$ being zero; i.e. 
the
state is invariant under $U(1)$.  

If however $a=b=0$ and $c$ and $d$ are non-zero
with $|c| \neq |d|$, then we find that invariance is further enhanced and the 
state
is invariant under $U(1)^2$.  If $|c| = |d|$, the state has yet further 
symmetry,
namely $U(1)\times SU(2)$  and one notices that the state is of
the form a singlet with respect to particles $1$ and $3$ tensor product with a 
vector
for particle 2; we write this as $\hbox{\em singlet}_{13}\otimes\hbox{\em 
vector}_2$.  The
invariance group $U(1)\times SU(2)$ arises since a singlet is invariant under a
(diagonal) $SU(2)$ and the state $\hbox{\em vector}_2$ is invariant under 
$U(1)$.  The
invariance group of the state cannot be increased by choosing special (non-zero)
values of $c$ and $d$ so a state of the form
$\hbox{\em singlet}_{13}\otimes\hbox{\em vector}_2$ has maximal symmetry. 

If $a=b=0$ and one
of $c$ or  $d$ are  also zero, we find the symmetry is also enhanced with 
respect to
the case where $c$ and $d$ are non-zero: in this case the symmetry is $U(1)^3$ 
and
such a state also has  maximal symmetry in the sense that no state has symmetry 
group
of which this is a subset.  The state is of the form $w_1\otimes w_2\otimes w_3$ 
(i.e. it is
homogeneous). In the case that $a=b=c=0$ the generators may be taken to be
$(\sigma_z)_1 + 1_8$,  $(\sigma_z)_2 - 1_8$ and $(\sigma_z)_3 - 1_8$, for 
example.

One also finds a similar structure among the states with $a=0$ and $c=0$ or 
$a=0$ and
$d=0$, namely invariance group of $U(1)^2$ in  unless the state is one of the 
special
ones with maximal symmetry namely either homogeneous with invariance $U(1)^3$, or 
of the
form $\hbox{\em singlet}\otimes\hbox{\em vector}$ with invariance $SU(2)\times 
U(1)$.

The cases of the sets of states with $b=0$ or $c=0$ have similar structure to 
those
with $a=0$.  The case of $d=0$ is different, however.

If $d=0$ and $a,b$ and $c$ are all non-zero, one calculates that the state is
annihilated by $(\sigma_z)_1 - (\sigma_z)_2 $ only; the state is invariant under
$U(1)$.  If  $d=0$ and $a=0$ but $b$ and $c$ are non-zero, the invariance is 
enhanced
to $U(1)^2$, in general or $SU(2)\times U(1)$ when $|b|=|c|$ in which case the
state is of the form $\hbox{\em singlet}_{12}\otimes\hbox{\em vector}_{3}$, a 
state of
maximal symmetry. When  $d=0$ and $b=0$ but $a$ and $c$ are non-zero, the 
invariance
is enhanced to $U(1)^3$; the state is homogeneous.  

Perhaps the most interesting case is when $d=0$ and $c=0$ but $a$ and $b$ are
non-zero, in which case one finds, for all values of $a$ and $b$, that the state 
is
invariant under $U(1)^2$.  However although there are a number of states with 
this symmetry, 
thought of as an abstract group, as described
above, the way that the group
acts on the states is quite different in the case $d=c=0$ than for example 
$d=a=0$.
In the case $d=c=0$, the generators are $(\sigma_z)_1 -(\sigma_z)_2$ and 
$(\sigma_z)_2 -(\sigma_z)_3$; corresponding to correlation between spins $1$ and 
$2$
and between $2$ and $3$.  In the case of $d=a=0$, the invariance group arises 
since
any vector in $\mathC^2$ is invariant under $U(1)$ and a generic two particle 
state
is also invariant under $U(1)$.

Amongst those states with $d=c=0$, there are some which larger symmetry groups 
than
$U(1)^2$.  If $a=0$ or $b=0$, then the invariance group is $U(1)^3$; the state 
is
homogeneous.  However the case $a=b$, while not having further continuous 
symmetry is
picked out by the fact that only this state has a discrete symmetry of $Z_2$
corresponding to the operation of simultaneously flipping all spins.  This is 
the
famous GHZ \cite{GHZ} state.

 \section{Conclusion}

In this paper we have started to build a general framework for understanding 
multi-particles entanglement. Obviously we have taken just a few steps here, and 
there are far more questions still open than answered. For example, it is known 
that in case of two-particle entanglement, to get a deeper understanding of 
entanglement one needs to take into account not only local unitary 
transformations but also measurements and classical communication between the 
two observers situated near the two particles. Also one has to consider actions 
taken on a large number of copies of the state $\Psi$ and not only on a single 
copy as considered here. Nevertheless, it is clear that any ``measure of 
entanglement" for multi-particles must be a function of the invariants described 
here.

 \bigskip 
\noindent{\large\bf Acknowledgments}

\noindent
We thank Graeme Segal and Tony Sudbery for very useful discussions.  We are also
very grateful to the Leverhulme and Newton Trusts for the financial support
given to NL.

\end{document}